\documentclass[prl,twocolumn,showpacs,preprintnumbers,amsmath,amssymb]{revtex4}
\usepackage[dvips]{graphicx}% Include figure files
\usepackage{dcolumn}% Align table columns on decimal point
\usepackage{bm}% bold math
\usepackage{times}
\usepackage{mathptmx}

\begin{document}

\title{Spin Oscillations in Antiferromagnetic NiO Triggered by
 Circularly Polarized Light}

\author{Takuya~Satoh$^1$}
\email{tsatoh@iis.u-tokyo.ac.jp}

\author{Sung-Jin~Cho$^1$}

\author{Ryugo~Iida$^1$}

\author{Tsutomu~Shimura$^1$}

\author{Kazuo~Kuroda$^1$}

\author{Hiroaki~Ueda$^2$}

\author{Yutaka~Ueda$^2$}

\author{B.~A.~Ivanov$^{3,4}$}

\author{Franco Nori$^{4,5}$}

\author{Manfred~Fiebig$^{6}$}

\affiliation{$^1$Institute of Industrial Science, University of
Tokyo, Tokyo 153-8904, Japan}

\affiliation{$^2$Institute for Solid State Physics, University of
Tokyo, Kashiwa, Chiba 277-8581, Japan}

\affiliation{$^3$Institute of Magnetism, Vernadskii Ave. 36B, 03142
Kiev, Ukraine}

\affiliation{$^{4}$RIKEN Advanced Science Institute, Wako-shi,
Saitama, 351-0198, Japan}

\affiliation{$^{5}$Department of Physics, The University of
Michigan, Ann Arbor, MI 48109-1040, USA, }

\affiliation{$^{6}$HISKP, Universit\"{a}t Bonn, Nussallee 14-16,
53115 Bonn, Germany}

\begin{abstract}
Coherent spin oscillations were non-thermally induced by circularly
polarized pulses in fully compensated antiferromagnetic NiO. This
effect is attributed to an entirely new mechanism of the action, on
the spins, of the effective magnetic field generated by an inverse
Faraday effect. The novelty of this mechanism is that spin
oscillations are driven by the time derivative of the effective
magnetic field acting even on ``pure'' antiferromagnets with zero
net magnetic moment in the ground state. The measured frequencies
(1.07~THz and 140~GHz) of the spin oscillations correspond to the
out-of-plane and in-plane modes of antiferromagnetic magnons.

\end{abstract}

 %magneto-optical effects

\date{
\today, Originally submitted on December 2, 2009}
%\hspace*{4cm}}

\maketitle

All-optical magnetization switching has been extensively studied in
recent years. In 1996, demagnetization within 1~ps was discovered by
irradiating ferromagnetic nickel with femtosecond laser pulses
\cite{beaurepaire96}. This pioneering finding has stimulated intense
theoretical and experimental investigations. Many of the experiments
on so-called ``ultrafast magnetism'' can be interpreted in terms of
laser-induced heating,
%The optical pump pulse transfers energy to
%the electrons, from where it is distributed to the lattice and the
%spins --- a process conveniently described by a three-temperature
%model.
which is already exploited technologically in
the form of heat-assisted magnetic recording \cite{Pan09}. However, this is
a relatively slow process since the recording rate is limited by
thermal diffusion. Thus, magnetization control beyond the limit of
such thermal control is highly desirable.

A typical form of non-thermal magnetization control is the inverse
Faraday effect (IFE). The IFE was predicted by Pitaevskii
\cite{pitaevskii61} and was demonstrated in non-absorbing media by
van der Ziel {\it et al.} \cite{ziel65}. Due to this effect,
circularly polarized light induces magnetization that can be
described as a light-induced effective magnetic field acting on the
body.
A pump--probe technique with sub-picosecond time resolution
has revealed transient IFE at zero time delay in itinerant
ferromagnets \cite{ju98,bennett99,wilks03,longa07}. However,
evidence of the spin-related contribution has been an issue because
no impact was observed after the temporal overlap of the pump and
probe pulses \cite{koopmans00}.

The dynamic properties of  antiferromagnets (AFMs) are rapidly
gaining importance \cite{kimel04,kimel05}. These compounds can
display inherently faster spin dynamics than ferromagnetic compounds
\cite{kimel04,kimel05,satoh07}, and offer the advantage that the
spin oscillation frequency extends into the subTHz and THz regime.
In addition, ultrafast manipulation of the antiferromagnetic order
parameter may be employed for ultrafast control of the magnetization
of an adjacent ferromagnet via the exchange-bias effect. Recently
spin precession caused by the IFE has been reported for
ferrite-garnets \cite{hansteen06} and for canted AFMs
\cite{kimel05,kalashnikova07}, which possess non-zero net magnetic
moment caused by the Dzyaloshinskii--Moriya interaction. The
presence of this magnetic moment is an important issue for the
recently proposed mechanism of inertia-driven excitation of spin
oscillations in canted AFMs \cite{kimel09}. On the other hand, a
fully non-thermal control of spin oscillations has not been
demonstrated yet in`pure'' AFMs,
%OLDOLD` having a fully compensated magnetic moment ($\vec M
%=0$).
%NEWNEW
having a fully compensated magnetic moment ($\vec M =0$) in the
ground state.

Here we report the first observation of coherent spin oscillations
in a fully compensated ($\vec M =0$) AFM NiO in a pump--probe
experiment. The oscillations consisted of 1.07~THz and 140~GHz
frequency components, which are assigned to out-of-plane and
in-plane modes of antiferromagnetic spin oscillations. The sign of
the oscillation was reversed with the reversal of the circularly
polarized pump helicity. This is interpreted within the
$\sigma$-model approach as a direct action of the \emph{time
derivative} of the impulsive magnetic field generated by a
circularly polarized pulse via the IFE on the zero-magnetization
AFM. This mechanism (discussed in Refs.
\onlinecite{andreev80,galkin08}, but never observed before) opens a
novel way for the ultrafast effective control of spins in
compensated AFMs.

NiO is one of the most promising exchange-bias AFMs because of its
simple structure and room-temperature antiferromagnetism. Therefore,
the investigation of the time-resolved responses of the electric,
magnetic, and optical properties of NiO could play an important role
for applications of ultrafast optical switching, and in
fundamental research. The sub-picosecond spin reorientation in NiO
has been demonstrated by modifying the magnetocrystalline anisotropy
with linearly polarized light~\cite{duong04}. However, this process
depends on a resonant optical excitation and is thus limited by
thermal effects.

%The optical and magnetic properties of NiO have been thoroughly studied.
Above the N\'eel temperature ($T_{\rm N}=523$~K), NiO has
an NaCl-type cubic structure (point group: $m\overline{3}m$).
Bolow $T_{\rm N}$, NiO has antiferromagnetic order. The Ni$^{2+}$ spins
align ferromagnetically along the $\langle 11\overline{2} \rangle$
axes in $\{111\}$ planes with antiferromagnetic coupling in between
adjacent $\{111\}$ planes \cite{hutchings72}. Exchange
striction leads to a contraction of the cubic unit cell along the
$\langle 111 \rangle$ axes and reduces the crystallographic symmetry
to $\overline{3}m$. This gives rise to four-types of twin (T)
domains. The deformation is accompanied by magnetic birefringence
between the $\{111\}$ plane and the $\langle 111 \rangle$ direction.
%with $\Delta n = n_{\rm e}-n_{\rm o} \simeq 0.003$ at $\lambda=$
%590~nm \cite{roth60}. Here, $n_{\rm e}$ and $n_{\rm o}$ are the
%refractive indices for the electric field components parallel and
%perpendicular to the $\langle 111 \rangle$ direction, respectively.
%A magnetostriction with a smaller distortion along the spin
%direction further reduces the crystallographic symmetry to $2/m$,
%which causes three-types of spin (S) domains for each T domain. The
%magnetic birefringence associated with this additional deformation
%is so small that it is usually negligible in linear optical
%experiments \cite{fiebig01}.
NiO is a charge-transfer insulator with a 4~eV band gap.  The
intra-gap optical transition in the mid-infrared to visible region
is ascribed to the electric-dipole forbidden $d$-$d$ transitions of
the Ni$^{2+}(3d^8)$ electrons. In particular, a $^3\Gamma_2 \to$
$^3\Gamma_4$ transition centered at 700~nm and a $^3\Gamma_2 \to$
$^3\Gamma_5$ transition centered at 1150~nm have been identified
\cite{fiebig01}.

\begin{figure}
\includegraphics[width=0.47\textwidth,clip]{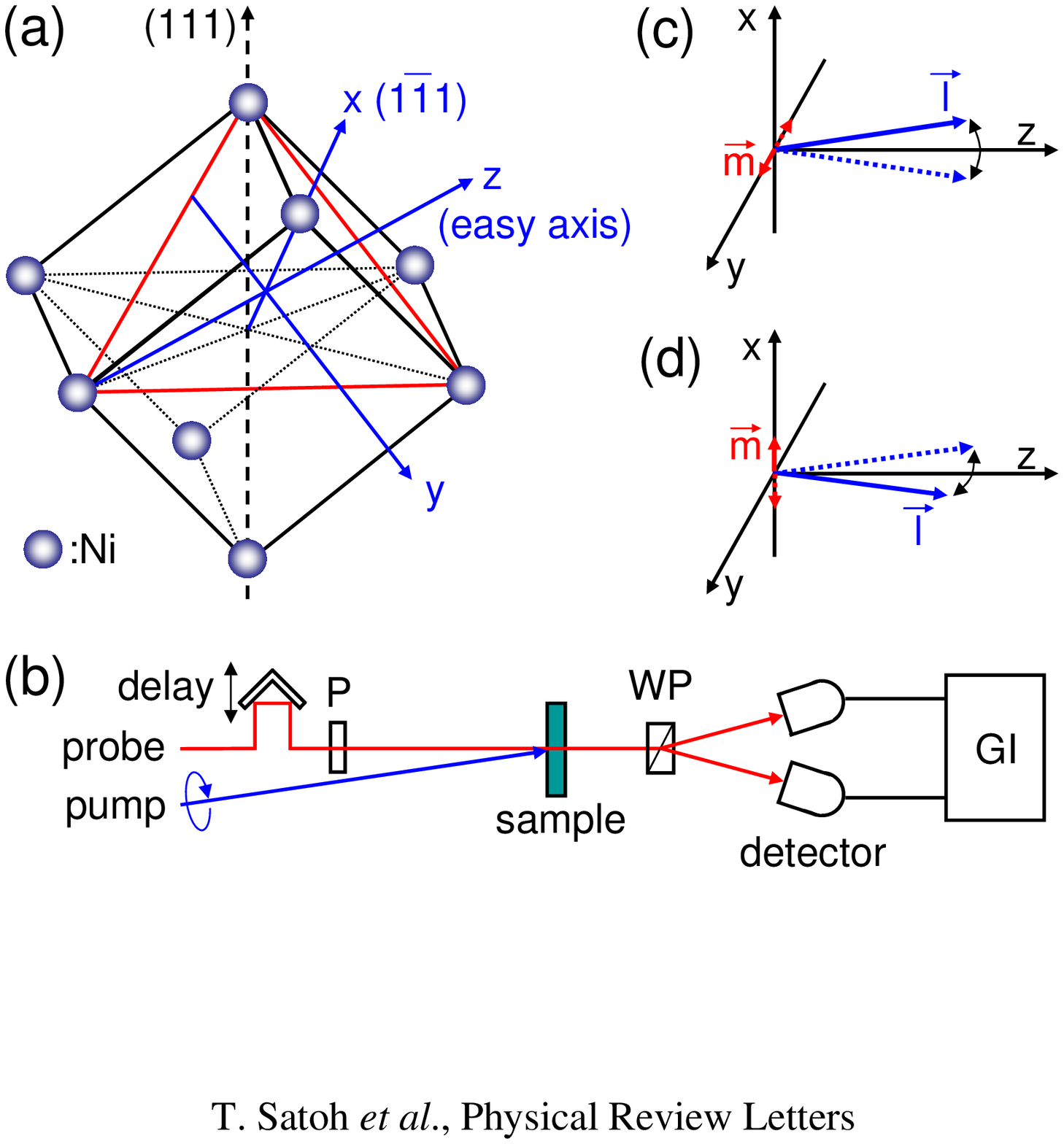}
\caption{\label{precession} (Color online)  (a) The geometry of the
experiment; the (111) direction coincides with the normal to the sample,
the $(1\bar11)$ direction ($x$-axis), $y$-axis, and $z$-axis,
respectively, are the hard, medium, and easy axes for the spins in
the T domain chosen for measurements (shown by the red triangle);
(b) Schematic diagram of the experimental setup. (P: polarizer, WP: Wollaston prism, GI: gated integrator);
(c), (d) show a schematics of the oscillations of vectors
$\vec l$ and $\vec m$ (blue and red arrows, respectively) for the
out-of-plane and in-plane modes of the spin oscillations in NiO.}
\end{figure}

A NiO single crystal was  grown by a floating-zone method. The bulk
sample was polished into (111)-oriented platelets with lateral
dimensions of a few millimeters and a thickness of $\simeq
100$~$\mu$m. As-grown samples possessed T domains with a lateral
size of $<1$~$\mu$m. To obtain T domains of 0.1--1~mm, the platelets
were annealed in an argon-oxygen mixture with small oxygen partial
pressure at 1400$^{\circ}$C \cite{saenger06}. By rotating a
polarizer and analyzer in the cross-Nicol configuration, four
types of T domains with a size of $\approx 500~\mu$m were
distinguished. For the pump--probe measurement, we selected a single
T domain with the $(1\overline{1}1)$ plane different from the sample
surface (111), as shown in Fig.~\ref{precession}(a).

The temporal evolution of the polarization rotation and
transmission were measured with a pump--probe setup (Fig.~\ref{precession}(b)).
The sample was in a cryostat at 77~K with no external magnetic field.
Linearly polarized light from a mode-locked Ti:sapphire laser with a wavelength of 792~nm, a
pulse width of 120~fs, and a repetition rate of 1~kHz was used as
the probe. Circularly polarized optical pulses with a wavelength of
1280~nm, generated by an optical parametric amplifier, were used as
the pump.
%Wave plates and polarizers were used to set the polarization of the incoming beams.
The pump and probe beams were focused on the sample
surface to spot sizes of about 100~$\mu$m and 40~$\mu$m,
respectively. The pump fluence was 10~mJ/cm$^2$, which corresponds
to the absorption of about one photon per $10^4$ Ni$^{2+}$
ions. The probe beam fell on the sample at normal incidence, whereas
the pump beam was incident at an angle of 7$^{\circ}$.
%The transmitted probe beam was detected after suppressing the residual pump beam with a color-glass filter.
The transmitted probe beam was divided into
two orthogonally polarized components by a Wollaston prism, and each
beam was detected with a Si photodiode to obtain the polarization rotation and the transmission change.
%Each signal was sent to a
%boxcar integrator connected to a computer. The polarization rotation
%and the transmission change were calculated by the computer.
%The dual detection technique allowed us to measure both values simultaneously.

%Figure 1

\begin{figure}
\includegraphics[width=0.47\textwidth,clip]{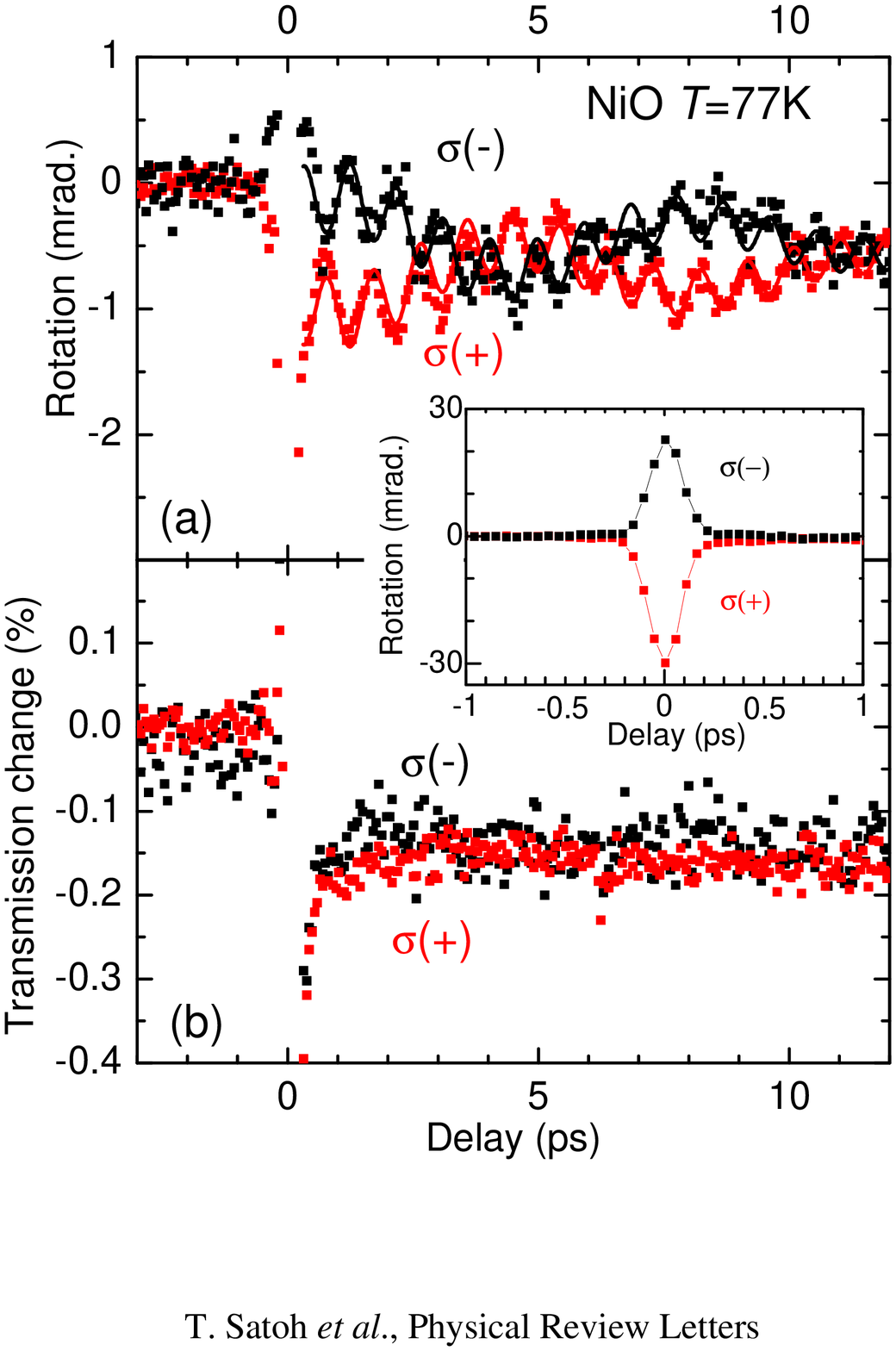}
\caption{\label{zero-delay} (Color online) Time-resolved (a)
polarization rotation (with a shorter time scale in the inset), and
(b) transmission change, in a NiO (111) sample for pump helicities
$\sigma_+$ and $\sigma_-$ with pump wavelength $\lambda_e$=1280~nm
and probe wavelength $\lambda_p$=792~nm.}
\end{figure}

To clarify the spin-related contribution,  we examined the
impact after photo-excitation with different time delays.
Figures~\ref{zero-delay}(a,b) show the polarization rotation and the
transmission change, respectively, versus the time delay between the
probe beam and the pump beam. The inset in Fig.~\ref{zero-delay}
shows the polarization rotation of the probe beam near zero-delay.
Here the signal is compared for $\sigma_+$ and $\sigma_-$ polarized
pump beams at fixed laser fluence. In Fig.~\ref{zero-delay}, two
processes can be distinguished: (1) a fast (practically
instantaneous) change of the polarization rotation within the time
of the pulse action (in the inset); and (2) damped oscillations of the
polarization rotation which persists for much longer times (upper
frame).

%Figure 2

In regime (1), for short time delays ($\lesssim 1$ ps), the rotation
exceeded 20 mrad when the pump and the probe beams overlapped
temporally. The full width at half-maximum of the signal was about
200~fs, which reflects the duration of the pump and probe pulses. In
regime (2), the slowly damped oscillations of the signal (with a
signal amplitude much lower than for short times) were observed at
times longer than 10 ps. For both time intervals, the sign of the
signal changes with reversal of the pump helicity, which is a clear
indication of the non-thermal origin of the effect. Note the
significant difference in the amplitude of the rotation angle
observed at these two time scales ($ \lesssim 0.2$~ps and $\geq 1$
ps) that likely reflects the difference of the mechanisms responsible
for them.

Process (1) can be considered as a typical example of so-called
\emph{femtomagnetic effects}, arising at times much shorter than the
thermalization time \cite{bigot09}. An adequate description of this
regime involves either a direct transfer of photon angular momentum
to the medium or a photo-enhanced transfer between orbital and spin
momenta \cite{hansteen06,longa07,zhang08,kurkin08,bigot09}. For our
compound, the ground state ($^3{}\Gamma_{2}$) of the Ni$^{2+}(3d^8)$
ion, the orbital momentum is quenched due to orbital non-degeneracy.
In the virtually excited state, the orbital momentum is $\pm 1$
depending on the helicity $\sigma_{\pm}$ of the pump beam, which
leads to the appearance of a transient magnetization. Recently,
\emph{ab initio} calculations~\cite{lefkidis09} of an ultrafast
laser-induced spin switch in NiO has demonstrated the possibility of
inducing a spin magnetic moment at tens of femtoseconds that results in
\emph{instantaneous} magneto-optical effects in this material.

In process (2), the sign of the oscillations in the polarization
rotation changes with reversal of the pump helicity, indicating that
the oscillation is triggered by non-thermal photo-excitations. No
oscillation is observed in the transmission, indicating that the
oscillation in the rotation is magnetic in origin. The damped
oscillations are fitted well with the form $\sum_{i=1,2}a_i
\cos(2\pi f_i t+\phi_i)\exp(-t/\tau_i)$, where $a$ is amplitude,
$f=\omega /2\pi $ is frequency, $\phi$ is phase, and $\tau$ is the
damping rate of the mode. The data in Fig.~\ref{zero-delay}(a) yield
$a_1=0.3$~mrad, $f_1=$1.07~THz, $\phi_1=1.0, \tau_1=15$~ps,
$a_2=0.375$~mrad, $f_2=$140~GHz, $\phi_2=2.2$, and $\tau_2=10$~ps.
It is clear that at these time scales some magnon modes are excited
in the spin system. The coherent spin oscillations modulate the
dielectric permittivity tensors,
%OLDOLD leading to a rotation of the probe polarization.
%NEWNEW
leading to a rotation of the probe polarization, see below for
details.
Since the net magnetization is
zero in the ground state of NiO, the mechanism inducing spin
oscillations is different from that in ferrimagnets or in canted
AFMs such as orthoferrites with the Dzyaloshinskyii--Moriya
interaction. Such excitation in NiO cannot be attributed to the
mechanism discussed in \cite{kimel09}, for which the presence of
\emph{non}-zero net magnetic moment in the ground state is
necessary. On the other hand, the driven dynamics of pure AFMs can
be described by the $\sigma$-model with the derivative $ d\vec
H(t)/dt$ as a driving force \cite{galkin08}. Let us discuss below
process (2) within the $\sigma$-model.

For the theoretical description of the spin excitations, we employ
a model of NiO with two sublattices with magnetizations $\vec M_1$
and $\vec M_2$, $|\vec M_1|=|\vec M_2|=M_0$, coupled by the
antiferromagnetic next-nearest neighbor exchange interaction $J$
\cite{hutchings72}. For such AFMs, the antiferromagnetic vector
$\vec L =\vec M_1-\vec M_2$ is the principal dynamical variable.
Within the $\sigma$-model, the equation for the normalized (unit)
antiferromagnetic vector $\vec l = \vec L /| \vec L|$ can be written
through the variation of the Lagrangian \cite{andreev80,sigma}
$\mathcal{L}[\vec l]=\mathcal{L}_0+\mathcal{L}_{\mathrm{int}}$,
where $\mathcal{L}_0$ describes the free oscillations of the spin
system:
\begin{equation}
\label{eq1} \mathcal{L}_0 = \frac{\hbar }{2 \gamma H_{\mathrm{ex}}
}\bigl(\frac{\partial \vec l}{\partial t}\bigr)^2 - w(\vec l ),\
w(\vec l )=g\mu_B(H_{a1} l_x^2 + H_{a2} l_y^2),
\end{equation}
$\mathcal{L}_{\mathrm{int}}$ determines the action of the light. The
magnetization $\vec M = \vec M_1+\vec M_2=2M_0\vec m $ is a slave
variable and can be written in terms of the vector $\vec l$ and its
time derivative:
\begin{equation}
\label{M} H_{\mathrm{ex}} \vec m = \bigr[\vec H - \vec l(\vec H
\cdot \vec l)\bigr] + \frac{1}{\gamma }\bigr(\frac{\partial \vec
l}{\partial t}\times \vec l\bigr)\;.
\end{equation}
Here, the first term determines the canting of the sublattices,
caused by the effective magnetic field $\vec H$, and the second term
describes the dynamic contribution \cite{andreev80,sigma}. The value
of the Lagrangian is presented per one spin, $\gamma = g\mu _B /
\hbar $ the gyromagnetic ratio, $g$ the Land\'e factor, $\mu _B $
the Bohr magneton,  $H_{\mathrm{ex}}=zSJ/g \mu _B$ the exchange
field of AFM, and $z=6$ is the number of next-nearest neighbors. For
NiO, $J$=221 K, which for $S=1$ gives $\gamma
H_{\mathrm{ex}}=zSJ/\hbar =$27.4 THz. We used the simplest form of
the biaxial anisotropy $w(\vec l )$, written in terms of the
out-of-plane anisotropy field $H_{a1}$ and much smaller in-plane
anisotropy field $H_{a2}$ \cite{hutchings72}.

Within the $\sigma$-model, the action of the circularly polarized
light can be described by an effective magnetic field $\vec H(t)
\propto (\vec E \times \vec E^*)$, corresponding to the IFE; in this
case
\begin{equation}
 \mathcal{L}_{\mathrm{int}}= -
\frac{\hbar }{H_{\mathrm{ex}}} \biggl(\vec H\cdot\bigl(\vec l\times
\frac{\partial \vec l}{\partial t}\bigr)\biggr).
\end{equation}

The variation of $\mathcal{L}[\vec l]$ gives the dynamical equations
for $\vec l$. In linear approximation over the deviation from the
ground state ($\vec l_{\mathrm{ground}}$ is parallel to the
$z$-axis, see Fig.~\ref{precession}) they read
\begin{equation}
\label{EqLin} \frac{\partial^2 l_x}{\partial t^2}+\omega^2_1
l_x=\gamma \frac{d H_y}{d t}; \ \frac{\partial^2 l_y}{\partial
t^2}+\omega^2_2 l_y=-\gamma \frac{d H_x}{d t},
\end{equation}
%+\frac{2}{\tau_1}\frac{\partial l_x}{\partial t}
%+\frac{2}{\tau_2}\frac{\partial l_y}{\partial t}
where $\omega_1=\gamma \sqrt{2H_{\mathrm{ex}}H_{a1}}$ and
$\omega_2=\gamma \sqrt{2H_{\mathrm{ex}}H_{a2}}$ are the frequencies
of the out-of-plane and in-plane antiferromagnetic spin
oscillations, respectively, and we omitted the dissipation terms.
For a short enough pulse ($\omega_{1,2}\Delta t \ll 1$),
Eq.~\eqref{EqLin} describes a quite universal behavior for AFM
\cite{galkin08}. Namely, after the pulse action, for times $t\gg
\Delta t$, the spin dynamics exhibits free oscillations with
frequencies $\omega_{1,2}$ and amplitudes $l_{x,y}(t=0) = a_{1,2}$
determined by the form of the pulse:
\begin{equation}
\label{ampl}
 a_1=\gamma \bar H_y\Delta t, \ a_2=-\gamma \bar H_x\Delta t, \
 \bar H_{x,y} \Delta t \equiv \int_{ - \infty }^{ + \infty }
{H_{x,y}(t)dt}.
\end{equation}
Thus, the amplitudes of the two components of the oscillations are
determined by the pulse field components, $H_y$ and $H_x$.
%NEWNEW
The oscillation of the vector $\vec {m}$ produces a modulation of
the antisymmetric part of the permittivity tensor
$\varepsilon _{ij} $, $\Delta \varepsilon _{ij}^a \propto e_{ijk} m_k $;
where $e_{ijk} $ is the absolute antisymmetric tensor. In linear
approximation, the out-of plane mode produces $\Delta \varepsilon _{xz}^a \propto m_y $, and
the in-plane mode produces $\Delta \varepsilon _{yz}^a \propto m_x $.

For our measurements, a T domain inclined to the surface of the
sample was chosen (see Fig.~\ref{precession}(a)), and the values of
$H_y$ and $H_x$ were approximately equal, which is in agreement with
the observation that $a_1 \sim a_2$. The measured frequency
$f_1=$1.07~THz is in good agreement with the out-of-plane mode of
the  antiferromagnetic spin oscillations. From the NiO out-of-plane
anisotropy field, $\gamma H_{a1}\simeq 23$ GHz \cite{hutchings72},
one obtains $f_1=$1.1~THz. The 1.07~THz component has been observed
in far-infrared antiferromagnetic resonance
\cite{kondoh60,sievers63} and Raman scattering \cite{lockwood92}.
Concerning the in-plane mode, the data for a small in-plane
anisotropy field $\gamma H_{a2}\sim 1$ GHz are not well
known~\cite{hutchings72}, but spin oscillations with a frequency
$f_2=$140~GHz have recently been observed in NiO using Brillouin
scattering \cite{milano04}. Therefore, this strongly suggests that
the observed oscillations in Fig.~\ref{zero-delay}(a) within the
wide time interval from 1 ps to tens of picoseconds
%OLDOLD
% are the spin oscillations along the easy axis triggered by the
% effective magnetic field $\vec H$ generated via the IFE, with the
% \emph{time derivative} of the field $d\vec H/dt$ working as a torque
% acting on the vector $\vec l$.
%NEWNEW
after the pulse is turned off, are spin oscillations around
their easy-axis, which are the usual spin-wave modes, triggered by
the effective magnetic field $\vec H$ generated via the IFE, with
the \emph{time derivative} of the field $d\vec H/dt$ working as a
torque acting on the vector $\vec l$.

The simultaneous observation of the magnetic response at both
short times and long times, in processes (1) and (2), allows us to
reach conclusions about the applicability of different approaches to
describe the spin dynamics. The Landau--Lifshitz equation for
ferromagnets (or, equivalently, the $\sigma$-model equation for AFMs)
describes the dynamics of spin systems in terms of only the
magnetization vector (or sublattice magnetizations, for AFMs). These
equations are valid for quasi-equilibrium states, where these
magnetizations are formed by the exchange interaction. This occurs
(according to our observations) for times corresponding to process
(2). On the other hand, our results show that the $\sigma$-model
approach (as well as any common theories treating the dynamics of
spin systems through the mean value of the magnetization) are not
sufficient to describe shorter times $t\lesssim  0.5$~ps, namely,
the behavior in regime (1). To stress this, it is enough to mention,
that the amplitudes of the oscillation, Eq.~\eqref{ampl}, do not contain
anisotropy fields, and it can be obtained by neglecting all
relativistic interactions in the AFM, except for the Zeeman
interaction of the spins with the pulsed magnetic field. In this
approximation within the $\sigma$-model approach, the projection of
the magnetization parallel to the field is conserved and cannot
appear during the action of a short ($\Delta t \ll 1/\omega_{1,2}$)
field pulse. For the $\sigma$-model, direct calculations show that
the static and dynamic contributions to the
magnetization, Eq.~\eqref{M}, compensate for each other in this short time
interval \cite{galkin08}. On the other hand, the signal observed at
these times is much higher than for process (2). Thus, to describe
the initial stage (1) of the process, one needs to use a more
detailed analysis involving the spin and orbital momenta of the
solid, as well as the angular momentum of photons \cite{lefkidis09}.

To conclude, the time-resolved magneto-optical response of
antiferromagnetic NiO provides a direct measure of magnetization
changes under the action of circularly polarized light. Remarkably,
we found that even compensated antiferromagnetic NiO shows spin
oscillations triggered non-thermally by a circularly polarized
pulse, with the time derivative $d\vec H/dt$ as the driving force
acting on the antiferromagnetic vector $\vec l$. The 1.07~THz and
140~GHz components are in good agreement with experimentally
reported frequencies of antiferromagnetic spin oscillations. The
results of our experiment show the possibility of extending the
potential of NiO as an antiferromagnetic constituent in spintronic
devices, from static and thermally-limited spin-dynamical
experiments into the promising range of all-optical magnetization
control at THz frequencies.

This work was supported by KAKENHI
%(19860020 and 20760008)
, SPP 1133 of the DFG, UNAS (BI), and the NSA/LPS/ARO and NSF (FN).

%Figure 1

%Figure 2

\end{document}